\documentclass[a4paper]{article}
\usepackage[T1]{fontenc}
\usepackage[utf8]{inputenc}
\usepackage[english]{babel}

\usepackage{graphicx}
\usepackage{hyperref}
\usepackage{authblk}
\usepackage{geometry}
\usepackage{setspace}
\usepackage{amsmath}
\usepackage{amssymb}
\usepackage{amsfonts}
\usepackage{braket}
\usepackage{multicol}

\newcommand{\keywordsenglish}{Keywords}

\renewenvironment{abstract}{%
        \begin{center}
	\begin{minipage}{14cm}
	{\textbf{\abstractname:}}
}{
        \end{minipage}
	\end{center}
}

\newenvironment{keywords}{
        \def\abstractname{\keywordsenglish}
	\begin{abstract}
}{
        \end{abstract}
}

\title{Algebraic solution for the classical harmonic oscillator}
\author{Murilo B. Alves\thanks{\scriptsize{Correspondence email address: murilo.alves@lnls.br}}}

\affil{\small{Brazilian Synchrotron Light Laboratory -- LNLS, \\
Brazilian Center for Research in Energy and Materials -- CNPEM, 13083-970, Campinas, SP, Brazil and \\ Gleb Wataghin Institute of Physics, University of Campinas -- UNICAMP, 13083-859, Campinas, SP, Brazil}}
\date{\today}
\begin{document}

\maketitle

\begin{abstract}
The harmonic oscillator is one of the most studied systems in Physics with a myriad of applications. One of the first problems solved in a Quantum Mechanics course is calculating the energy spectrum of the simple harmonic oscillator with analytic and algebraic approaches. In the algebraic solution, creation and annihilation operators are introduced to factorize the Hamiltonian. This work presents an algebraic solution for the simple harmonic oscillator in the context of Classical Mechanics, exploring the Hamiltonian formalism. In this solution, similarities between the canonical coordinates in a convenient basis for the classical problem and the corresponding operators in Quantum Mechanics are highlighted. Moreover, the presented algebraic solution provides a straightforward procedure for the quantization of the classical harmonic oscillator, motivating and justifying some operator definitions commonly used to solve the correspondent problem in Quantum Mechanics.
\end{abstract}


\vspace{6pt}

\begin{multicols}{2}
\section{Introduction}
Over the undergraduate curriculum in Physics, the expressions for the creation (raising) and annihilation (lowering) operators are commonly presented to the student for the first time during a Quantum Mechanics course. The operators are typically introduced to develop the algebraic solution of the harmonic oscillator. Later in the studies of Physics, these operators are extensively used, for example, in the second quantization formalism. Following this study timeline, it is easy to create an incorrect opinion that this type of algebraic structure is exclusive to the quantum context.

In Dirac's textbook on Quantum Mechanics~\cite{Dirac1947}, a solution of the quantum harmonic oscillator is developed by using a transformation of coordinates which, as Dirac mentions, is motivated from Classical Mechanics. However, Dirac did not discuss the origin of this transformation in Classical Mechanics nor reference where it can be found. A brief review about the history of the simple harmonic oscillator in Quantum Mechanics and the introduction of ladder operators to algebraically solve the problem can be found in~\cite{Rushka2020}.

The main goal of this work is to present an algebraic solution for the classical harmonic oscillator that should be easy to follow by any undergraduate student in Physics satisfying the prerequisites for a Quantum Mechanics course. We will then discuss how this classical solution can be adapted to the quantum version of the problem.

\section{Algebraic solution in Classical Mechanics}
In Classical Mechanics, the Hamiltonian function for the simple harmonic oscillator is given by:
\begin{equation}
    H = \frac{p^2}{2m} + \frac{m\omega^2x^2}{2},
    \label{eq:hamilton}
\end{equation}
where the one-dimensional case will be solved, for convenience. The variable $x$ is the position of the oscillator with respect to its equilibrium (in this case, the coordinate system was chosen so the equilibrium position is $x=0$), and $p$ is the momentum $p = mv$, $m$ is the oscillator mass and $v$ the speed. The quadratic potential $V(x) = m\omega^2 x^2/2$ is related to a linear restoring force $F(x)=-m\omega^2x$, by $F = -\frac{\partial V}{\partial x}$.

The proposal of the Hamiltonian formalism is to transform $n$ second order ordinary differential equations (ODEs) (obtained by Newton's second law) into $2n$ first order ODEs, where $n$ is the number of degrees of freedom of the system.

Let $(x, p)$ be the canonical variables and $n=1$. With the Hamiltonian function defined in Eq.~\eqref{eq:hamilton}, the Hamilton equations for the simple harmonic oscillator are~\cite{Goldstein2002}:
\begin{subequations}
\begin{align}
    \dot{x} & = \frac{\partial H}{\partial p} =\frac{p}{m},
    \label{eq:eqm_x}\\
    \dot{p} & = -\frac{\partial H}{\partial x} = -m\omega^2 x.
    \label{eq:eqm_p}
\end{align}
\end{subequations}

After these equations are presented, is quite common that $\ddot{x} = \dot{p}/m$ is calculated by taking the time derivative of Eq.~\eqref{eq:eqm_x} and then get $\dot{p}$ from Eq.~\eqref{eq:eqm_p} to obtain $\ddot{x} + \omega^{2}x = 0$, showing the equivalence with the Newtonian formalism. However, this process returns to $n$ second order ODEs, evading the original proposal of the Hamiltonian formalism.

We note that, since the harmonic oscillator potential is quadratic, Hamilton equations become a linear system of differential equations, which can be rewritten in the matrix form:
\begin{equation}
\frac{d}{dt}
\begin{bmatrix}
    x  \\
    p \\
\end{bmatrix}
 =
\begin{bmatrix}
    0 & 1/m \\
    -m\omega^2 & 0 \\
\end{bmatrix}
\begin{bmatrix}
    x  \\
    p \\
\end{bmatrix}.
\end{equation}

Note that we obtained a coupled system of ODEs. We can write the system more compactly with $\frac{d\vec{\eta}}{dt} = \text{M} \vec{\eta}$. where the phase-space vector $\vec{\eta} = (x, p)^{\intercal}$ is written at a given base $\left\{\hat{e}_x, \hat{e}_p\right\}$ of the phase space and M is the matrix representation of the Hamiltonian time evolution in this base. In Appendix~\ref{appendix:symplectic} there is a very brief discussion on the symplectic formalism.

To decouple the system, we must diagonalize the matrix M and solve the problem on the basis of eigenvectors. Solving for $\det(\text{M} - \lambda \text{I}) = 0$, we get the eigenvalues $\lambda_{\pm} = \pm i \omega$, where $i^2 = -1$. The corresponding eigenvectors are
\begin{equation}
    \vec{v}_{\pm} = \alpha \begin{bmatrix}
    1  \\
    \pm i m\omega \\
\end{bmatrix}.
\end{equation}

We should calculate how the components of the vector $\vec{\eta}$ is written in the basis of eigenvectors, where M is known to be diagonal and the problem is decoupled. With the components of $\vec{v}_\pm$ written in the basis $\left\{\hat{e}_x, \hat{e}_p\right\}$, we can calculate the change of basis matrix and its inverse.
\begin{subequations}
    \begin{align}
        \text{T} &=
        \alpha \begin{bmatrix}
            1 & 1  \\
            i m\omega & - i m\omega \\
        \end{bmatrix}, \\
        \text{T}^{-1} &=
        \frac{1}{2\alpha} \begin{bmatrix}
            1 & -\frac{i}{m\omega}  \\
            1 & +\frac{i}{m\omega} \\
        \end{bmatrix}.
    \end{align}
\end{subequations}

To change from the original canonical basis to the eigenvectors basis, the matrix $\text{T}^{-1}$ must be applied to $\vec{\eta}$, which is written on this basis as $\vec{a}$:
\begin{equation}
    \vec{a} = \text{T}^{-1}\vec{\eta} = \frac{1}{2\alpha} \begin{bmatrix}
    x - \frac{ip}{m\omega} \\
    x + \frac{ip}{m\omega}  \\
\end{bmatrix}.
\label{eq:a_vec}
\end{equation}

Writing $\vec{a} = (a_{-}, a_{+})^{\intercal}$, note that $a_{-} = a^{*}_{+}$, where the symbol $^{*}$ denotes the complex conjugate. Then, let us define $a := a_{+}$ and $a^* := a_{-}$.

From Eq.~\eqref{eq:a_vec} we obtain that
\[
2 \alpha^{2}a^*a = \frac{p^2}{2m^{2}\omega^{2}} + \frac{x^2}{2} = \frac{H}{m\omega^2}.
\]

If we let the constant $\alpha$ be $
\alpha~=~1\slash \sqrt{2m\omega}$ and define $J = a^{*}a$, then $H = \omega J$ follows. This expression for the Hamiltonian can be achieved through canonical transformation to the well-known action-angle variables $(J, \phi)$. In these variables, the Hamilton equations are
\begin{subequations}
\begin{align}
    \dot{\phi} & = \frac{\partial H}{\partial J} = \omega,\\
    \dot{J} & = -\frac{\partial H}{\partial \phi} = 0.
\end{align}
\end{subequations}

This shows that the action variable $J=a^{*}a$, is a constant of motion.

Given two functions of the canonical variables, $f=f(x, p)$ and $g=g(x, p)$, the Poisson brackets between the functions are defined as~\cite{Goldstein2002}:
\begin{equation}
    \left\{f, g\right\} := \frac{\partial f}{\partial x}\frac{\partial g}{\partial p} - \frac{\partial f}{\partial p}\frac{\partial g}{\partial x}.
\end{equation}

A transformation $(x, p) \to (X, P)$ is called canonical in Classical Mechanics if $\left\{X, P\right\} = 1$ follows. Calculating the Poisson brackets of the new variables yields to $\left\{a, a^{*}\right\} = -i$. Therefore, with this specific choice for the constant $\alpha$, the transformation $(x, p) \to (a, a^{*})$ is not canonical.

With $\alpha = 1\slash \sqrt{2m\omega}$ as defined above, the components of $\vec{\eta}$ in the eigenvectors basis are given by:
\begin{subequations}
\begin{align}
    a & = \sqrt{\frac{m\omega}{2}}\left(x + i \frac{p}{m\omega}\right), \label{variable_a} \\
    a^{*} & = \sqrt{\frac{m\omega}{2}}\left(x - i \frac{p}{m\omega}\right) \label{variable_a*}.
\end{align}
\end{subequations}

We calculate the time evolution of the functions $a, a^*$ via Poisson brackets with the Hamiltonian:
\[
\dot{a} = \left\{a, H\right\} = - i\omega a, \,\,\,\, \dot{a}^* = \left\{a^*, H\right\} = +i\omega a^*,
\]
where $H = \omega a^{*}a$ was considered. As intended, the system on this basis is decoupled:
\begin{equation}
\frac{d}{dt}
\begin{bmatrix}
    a^*  \\
    a \
\end{bmatrix}
 =
\begin{bmatrix}
    +i\omega & 0 \\
    0 & -i\omega \\
\end{bmatrix}
\begin{bmatrix}
    a^*  \\
    a \\
\end{bmatrix}.
\label{eq:decoupled}
\end{equation}

In that form, the differential equations can be solved by simple integration from $t_0$ to $t$ and the result is
\[a(t) = a(t_0)e^{-i\omega (t-t_0)}, \,\,\,\, a^{*}(t) = a^*(t_0)e^{+i\omega (t-t_0)}.
\]
Note that $a^{*}(t)a(t) = a^{*}(t_0)a(t_0)$, explicitly showing that $a^*a$ is a constant of motion.

The variables $x(t)$ and $p(t)$ can be written in terms of $a(t)$ and $a^*(t)$:
\begin{subequations}
    \begin{align}
        x & = \sqrt{\frac{1}{2m\omega}}\left(a^{*} + a\right) = \sqrt{\frac{2}{m\omega}}\mathrm{Re}(a), \label{x_interms_a}\\
        p & = i\sqrt{\frac{m\omega}{2}}\left(a^{*} - a\right) = \sqrt{2m\omega}\mathrm{Im}(a). \label{p_interms_a}
    \end{align}
\end{subequations}

Without loss of generality, let us set the initial time as $t_0=0$ and the corresponding initial conditions to be $x(0) = x_0$ and $p(0) = p_0$. Thus, from Eq.~\eqref{variable_a}, the initial condition for the complex variable $a(t)$ is:
\[
a(t_0) = a_0 =  \sqrt{\frac{m\omega}{2}}\left(x_0 + i\frac{p_0}{m\omega}\right).
\]

Note that, since $a(t)$ is complex, to determine its initial value, two real numbers must be given, which can be related to two initial conditions of position and momentum that are commonly used. Thus, after all, two initial conditions are still required to determine the solution, since we are solving a second-order ODE. We showed that the time dependence of $a(t)$ is simply $a(t) = a_0e^{-i\omega t}$. Then, with Euler's identity and the expression for $a_0$, we obtain:
\begin{align}
    \mathrm{Re}\left[a(t)\right] &= \sqrt{\frac{m\omega}{2}}\left[x_0 \cos(\omega t) + \frac{p_0}{m\omega}\sin(\omega t)\right] \\
    \mathrm{Im}\left[a(t)\right] &= \sqrt{\frac{m\omega}{2}}\left[\frac{p_0}{m\omega} \cos(\omega t) - x_0\sin(\omega t)\right]
\end{align}

Finally, applying this result to Eqs.~\eqref{x_interms_a} and~\eqref{p_interms_a}, we reach to the well-known solution of the simple harmonic oscillator.
\begin{subequations}
    \begin{align}
        x(t) &= x_0 \cos(\omega t) + \frac{p_0}{m\omega}\sin(\omega t)
        \label{eq:solutionx}, \\
        p(t) &= p_0 \cos(\omega t) - m\omega x_0 \sin(\omega t)
        \label{eq:solutionp},
    \end{align}
\end{subequations}
Note that Eq.~\eqref{eq:solutionx} and~\eqref{eq:solutionp} can be written in a matrix form
\begin{equation*}
\begin{bmatrix}
    x(t)  \\
    p(t) \\
\end{bmatrix}
 =
\begin{bmatrix}
    \cos(\omega t) & \sin(\omega t)/m\omega \\
    -m\omega\sin(\omega t) & \cos(\omega t) \\
\end{bmatrix}
\begin{bmatrix}
    x_0  \\
    p_0 \\
\end{bmatrix}.
\end{equation*}
In this form, the time evolution of the harmonic oscillator can be interpreted as a time-dependent transfer matrix $\text{U}_{t \leftarrow t_0}$ that propagates the variables $x$ and $p$ from $t_0$ to any $t$. Therefore, the phase-space vector $\vec{\eta}(t) = (x(t), p(t))^{\intercal}$ is obtained from the initial condition $\vec{\eta}(t_0) = (x_0, p_0)$ with $\vec{\eta}(t) = \text{U}_{t \leftarrow t_0} \vec{\eta}(t_0)$.

\section{Quantization}
\subsection{Bohr-Sommerfeld}
As presented on the previous section, $a^*a$ is an invariant. Explicitly calculating this product, we have
\begin{equation}
    a^*a = \frac{m\omega}{2}\left(x^2 + \frac{p^2}{m^{2}\omega^{2}} + \frac{i}{m\omega}(xp - px)\right)
    \label{eq:commute}
\end{equation}
where the term $(xp - px)$ is zero in Classical Mechanics but it was intentionally kept on Eq.~\eqref{eq:commute} to highlight the relevance of commutativity between products of variables on the derivation.

We observed that the Hamiltonian function can be written simply as $H = \omega a^{*}a = \omega J$. A first attempt to study the quantum analog of the classical harmonic oscillator consists in applying the Bohr-Sommerfeld quantization rule, imposing the condition $J = n\hbar$, where $n$ is an integer. Thus, the energy levels of the system, obtained by the Hamiltonian, are $E_n = n\hbar \omega$. Note that in this case the ground state energy obtained is $E_0=0$, which is currently known to be incorrect. Nevertheless, this is a first step on the path of quantization of energies and its relation to the oscillation frequencies. Furthermore, we see that $J = a^{*}a = n\hbar$, which is an indicative that the product of the variables $a^*a$ should be related to an integer number $n$ that can be associated to the oscillator energy level.

\subsection{Dirac's correspondence}
Another attempt, currently known to be the correct one, is applying the correspondence principle proposed by Dirac for the quantization of classical systems~\cite{Dirac1947}, also known as canonical quantization. In this procedure the canonical variables and their functions become operators in a Hilbert space and the Poisson brackets are replaced by commutators of operators $\left\{\cdot, \cdot \right\} \to \frac{1}{i\hbar}\left[\cdot, \cdot \right]$, where $\left[\hat{f}, \hat{g} \right] = \hat{f}\hat{g} - \hat{g}\hat{f}$. From this point of the work, operators will be denoted with hats.

The first observation is that the classical relation between the variables $(a, a^{*})$ given by $\left\{a, a^{*}\right\} = -i$, becomes $\left[\hat {a}, \hat{a}^{\dagger}\right] = \hbar$ with this correspondence. The symbol $^{\dagger}$ denotes the Hermitian adjoint of an operator. To make the commutation relation be $\left[\hat{a}, \hat{a}^{\dagger}\right] = 1$, as is typically defined in the literature, we can divide by a factor $\sqrt{\hbar}$ in each one of the operators $\hat{a}, \hat{a}^{\dagger}$, thus obtaining:
\begin{subequations}
    \begin{align}
        \hat{a} & = \sqrt{\frac{m\omega}{2\hbar}}\left(\hat{x} + i \frac{\hat{p}}{m\omega}\right), \\
        \hat{a}^{\dagger} & = \sqrt{\frac{m\omega}{2\hbar}}\left(\hat{x} - i \frac{\hat{p}}{m\omega}\right).
    \end{align}
\end{subequations}

These are the creation and annihilation operators that are commonly introduced in a Quantum Mechanics course~\cite{Sakurai2014}, often without prior motivation, during the development of the algebraic solution of the quantum harmonic oscillator. The number operator is also commonly defined~\textit{ad hoc} as $\hat{N} = \hat{a}^{\dagger}\hat{a}$. With our approach, the reason for those definitions can be justified.

With Dirac's correspondence, the fundamental relation $\left\{x, p\right\} = 1$ translates to $\left[\hat{x}, \hat{p}\right] = i\hbar$. Then, the explicit calculation of the product $\hat{a}^{\dagger}\hat{a}$, as done in Eq.~\eqref{eq:commute}, results in
\begin{align*}
    \hat{a}^{\dagger}\hat{a} &= \frac{\hat{H}}{\hbar \omega} + \frac{i}{2\hbar}\left[\hat{x}, \hat{p}\right] \\
    &= \frac{H}{\hbar\omega} - \frac{1}{2}.
\end{align*}

Then, the Hamiltonian operator is
\begin{equation}
    \hat{H} = \hbar \omega (\hat{a}^{\dagger}\hat{a} + 1/2).
\end{equation}

Let a basis of autokets be such that $\hat{N}\ket{n} = n \ket{n}$ and the operator number $\hat{N}$ defined as $\hat{N} = \hat{a}^{\dagger}\hat{a}$. Thus, $\hat{H} = \hbar \omega (\hat{N}+ 1/2) $. Note that the energy levels are $E_n = \hbar\omega (n + 1/2)$ and the ground state energy is $E_0 = \hbar\omega /2 \neq 0$. It also becomes explicit that the non-zero energy of the ground state of the harmonic oscillator is related to the fact that the product between the operators $\hat{x}$ and $\hat{p}$ is non-commutative, which in turn is closely related to the Heisenberg's uncertainty principle.

Following Heisenberg description~\cite{Sakurai2014}, time evolution takes place on quantum operators instead of quantum states, according to the Heisenberg equation:
\begin{equation}
\frac{d\hat{A}}{dt} = \frac{1}{i\hbar}\left[\hat{A}, \hat{H}\right],
\end{equation}
for an operator $\hat{A}$. In this description, we can follow exactly the same steps used in the calculation of the classical time evolution of the variables $a$ and $a^{*}$, in order to calculate the evolution of the operators $\hat{a}$ and $\hat{a} ^{\dagger}$ and finally obtain solutions with expressions identical to Eqs.~~\eqref{eq:solutionx} and~\eqref{eq:solutionp}, but with the operators position $\hat{x}$ and momentum $\hat{p}$.

\section{Conclusion}
We presented an algebraic solution for the classical harmonic oscillator exploring the Hamiltonian formalism. In this solution the components of the position-momentum vector in the phase space written in the basis that diagonalizes the Hamilton equations have an expression very similar to the creation (raising) and annihilation (lowering) quantum operators. It was possible to develop a motivation from Classical Mechanics for the origin of these operators in Quantum Mechanics and also justify the number operator definition. Using the correspondence principle for quantization of classical systems proposed by Dirac, the algebraic solution to the quantum problem is obtained as a natural consequence of the classical solution. Furthermore, it was possible to explicitly highlight one of the differences between a classical and a quantum theory manifested in the ground state energy of the harmonic oscillator, which is directly related to the commutative property of the variables in the classical theory in contrast to the non-commutativity of the corresponding operators of the quantum theory.

\section*{Acknowledgment}
The author thanks F. H. de Sá for valuable suggestions on the manuscript.

\appendix
\section*{Appendix A - Symplectic notation}\label{appendix:symplectic}
Using the symplectic notation the Hamilton equations can be written as~\cite{Aguiar2011}:
\begin{equation}
    \frac{\mathrm{d}\vec{\eta}}{\mathrm{d}t} = \text{J}\nabla_{\vec{\eta}}H,
\end{equation}
where $\vec{\eta} = (x, p)^{\intercal}$ e $\nabla_{\vec{\eta}} = \left(\frac{\partial}{\partial x}, \frac{\partial}{\partial p}\right)^{\intercal}$.

The matrix J in the one-dimensional case has the expression
\begin{equation}
    \text{J} = \begin{bmatrix}
    0 & 1 \\
    -1 & 0 \\
\end{bmatrix},
\end{equation}
and it is called \textit{fundamental symplectic matrix}. The description can be generalized to $N$ dimensions and the matrix J satisfy the relations
\[
\text{J}^2 = -\text{I}, \,\,\,\,\, \text{J}^\intercal = -\text{J},
\]
from which $\text{J}^\intercal \text{J} = \text{I}$ follows.

A matrix $\Omega$ is symplectic if satisfies $\text{J}^\intercal \Omega \text{J} = \Omega$. The term $\text{J}\nabla_{\vec{\eta}}$ is also known as \textit{symplectic gradient}.

The harmonic oscillator time evolution matrix $M$ is symplectic. If the coordinate transformation $P = p/\sqrt{m}$ and $X = \omega \sqrt{m} x$ is applied, the Hamiltonian function is written as $H = (P^2 + X^{2})/2$ and the matrix $M$ in these coordinates is equal to the fundamental symplectic matrix J.

\section*{Appendix B - Canonical transformation}\label{appendix:canonical}
In the third edition of Goldstein's textbook of Classical Mechanics~\cite{Goldstein2002}, there is an exercise in chapter about Canonical Transformation (Chapter 9) that asks the reader to prove that the transformation
\begin{equation*}
 X = p + iax, \,\,\,\,\,  P = \frac{p - iax}{2ia},
\end{equation*}
is canonical, where $a$ is a constant. Then, the exercise suggests to use this canonical transformation to solve the linear harmonic oscillator.

With Poisson brackets, it can be shown that $\left\{X, P\right\} = 1$ then the transformation is canonical ($\left\{X, X\right\} = \left\{P, P\right\} = 0$ are obvious). Furthermore, $2iaXP = p^2 + (a x)^2$ readily follows. From the Hamiltonian function for harmonic oscillator, we note that $2mH = p^2 + (m\omega x)^2$, then the choice of constant $a = m\omega$ is quite natural. With $a = m\omega$, the two expressions can be compared to obtain the new Hamiltonian after this canonical transformation as $H' = i\omega XP$. Thus, the Hamilton equations are
\begin{subequations}
    \begin{align}
        \dot{X} & = \frac{\partial H'}{\partial P} = i\omega X,\\
        \dot{P} & = -\frac{\partial H'}{\partial X} = -i\omega P,
    \end{align}
\end{subequations}
which can be compared to Eq.~\eqref{eq:decoupled}, by identifying the similarities $X \sim a^*$ and $P \sim a$. The time evolution of $X(t)$ and $P(t)$ can be obtained by simple integration as $X(t) = X(t_0)e^{i\omega t}$ and $P(t) = P(t_0)e^{-i\omega t}$ as well. By inverting the canonical transformation to write $x(X, P)$ and $p(X, P)$, one obtains the oscillatory solutions as in Eqs.~\eqref{eq:solutionx} and~\eqref{eq:solutionp}.

The transformation proposed by Goldstein's exercise is canonical, since $\left\{X, P\right\} = 1$ is verified. With this canonical transformation the new Hamiltonian is complex: $H = i\omega XP$. The transformation presented in this paper is not canonical, since $\left\{a, a^*\right\} = -i$, but the corresponding Hamiltonian is real, given by $H = \omega a^*a$.

\end{multicols}
\end{document}